\input epsf
\input rotate
\documentstyle[bolditalic,prb,aps,12pt]{revtex}

\begin{document}

\title{Two-fluid magnetic island dynamics in slab geometry:\\
I - Isolated islands}
\author{~\\Richard Fitzpatrick\thanks{rfitzp@farside.ph.utexas.edu}
and Fran\c{c}ois L.\ Waelbroeck\\
{\em Institute for Fusion Studies}\\
{\em Department of Physics}\\
{\em University of Texas at Austin}\\
{\em Austin, TX 78712}\\
~\\}
\maketitle
\begin{abstract} 
A set of reduced, 2-D, two-fluid, drift-MHD (magnetohydrodynamical)
equations is derived. Using these equations, a complete and fully
self-consistent solution is obtained for an isolated magnetic island
propagating through a slab plasma with uniform but different
ion and electron fluid velocities. The ion and electron fluid flow profiles
around the island are uniquely determined, and are everywhere
continuous. Moreover, the island phase-velocity is uniquely specified by the
condition that there be zero net electromagnetic
force acting on the island. Finally, the ion polarization current correction
to the Rutherford island width evolution equation is evaluated, and
found to be stabilizing provided that the anomalous perpendicular
ion viscosity significantly exceeds the anomalous perpendicular
electron viscosity.

\end{abstract}
\section{Introduction}
Tearing modes are magnetohydrodynamical (MHD) instabilities
which often limit fusion plasma performance in magnetic 
confinement devices relying on nested toroidal magnetic
 flux-surfaces.\cite{ros} As the name suggests, ``tearing'' modes tear and reconnect magnetic
field-lines, in the process converting  nested toroidal flux-surfaces into helical magnetic islands.
Such islands degrade plasma confinement because
heat and particles are able to travel radially from one side of an island
to another by flowing along magnetic field-lines, which is a
relatively fast process, instead of having to diffuse across
magnetic flux-surfaces, which is a relatively slow process.\cite{callen}

Magnetic island physics is very well understood within the
context of {\em single-fluid}\/ MHD theory.  According to this
theory, the island width
is governed by the well-known  nonlinear evolution equation due to
Rutherford.\cite{ruth} Moreover, the island is required to propagate at the
local flow velocity of the MHD fluid, since fluid flow across
the island separatrix is effectively prohibited.

Magnetic island physics is  less completely
understood within the context of {\em two-fluid}, drift-MHD 
theory,\cite{gar1,smol,gar2,reb,zab,con,smol1,wil,wal,kush,rf,mik,con1,shaing,fr}
which is far more relevant to present-day magnetic confinement devices than
single-fluid theory.  In two-fluid theory, the island is generally embedded
within ion and electron fluids which flow at {\em different} velocities.
The island itself usually propagates at some intermediate velocity.
For sufficiently wide islands, both fluids are required to flow at the island propagation velocity
in the region lying within the island separatrix (since neither fluid can easily
cross the separatrix). However, the region immediately
outside the separatrix is characterized by strongly sheared ion
and electron fluid flow profiles, as the velocities of both fluids
adjust to their unperturbed values far away from the island.
The polarization current generated by the strongly sheared
ion  flow around the island separatrix gives rise to an additional term in the Rutherford
island width evolution equation, which is stabilizing or
destabilizing, depending on the island propagation velocity relative
to the unperturbed flow velocities of the ion and MHD fluids. The key 
problems in two-fluid island theory are the unambiguous determination of
the island phase-velocity, and the calculation of
the ion and electron fluid flow
profiles around the island separatrix. As yet, no consensus
has emerged within the magnetic fusion community regarding  the
solution of these problems.

In this paper, we first develop a set of reduced, 2-D, two-fluid,
drift-MHD equations. These equations contain both electron and ion
diamagnetic effects (including the contribution of the ion gyroviscous
tensor), as well as the Hall effect and parallel electron compressibility.
However, they do not contain electron inertia or the compressible Alfv\'{e}n wave
(which play  negligible roles in conventional magnetic island
physics). Our set of equations
consist of four coupled partial differential equations, and is both
analytically tractable and easy to solve numerically. 
We  employ our equations to study
the evolution of an isolated magnetic island in  {\em slab geometry}. Using a particular
ordering scheme, we are able to calculate the island phase-velocity, and to uniquely determine the ion and electron fluid flow
profiles outside the island separatrix.

\section{Derivation of reduced equations}
\subsection{Introduction}
In this section, we shall generalize the analysis of Refs.~\onlinecite{rf1}
and \onlinecite{franco} to obtain a set of reduced, 2-D,
two-fluid,
drift-MHD
equations which take ion diamagnetic flows into account.

\subsection{Basic equations}
Standard right-handed Cartesian coordinates ($x$, $y$, $z$) are adopted.
Consider a quasi-neutral plasma with singly-charged ions. The 
ion/electron number density $n_0$ is assumed to be {\em uniform} and {\em constant}.
Suppose that $T_i = \tau\,T_e$, where $T_{i,e}$ is the ion/electron
temperature, and $\tau$ is {\em uniform} and {\em constant}. 

Broadly following Ref.~\onlinecite{haz}, we adopt the following set of
two-fluid, drift-MHD equations:
\begin{eqnarray}
{\bfm E} + {\bfm V}\times{\bfm B} +\frac{1}{e\,n_0}\!
\left(\nabla \,P - \frac{\tau}{1+\tau}\,({\bfm b}\cdot\nabla P)\,{\bfm b}
-{\bfm J}\times{\bfm B} - \mu_e\,\nabla^2{\bfm V}_e\right)
&=&\eta\,\!\left({\bfm J} - \frac{3}{2}\,\frac{\tau}{1+\tau}\,n_0\,e\,
{\bfm V}_\ast\right),\nonumber\\[0.5ex]
&&\\[0.5ex]
m_i\,n_0\!\left[\left(
\frac{\partial}{\partial t} + {\bfm V}\!\cdot\!\nabla + \frac{\tau}{1+\tau}\,
{\bfm V}_\ast\!\cdot\!\nabla\right)\!{\bfm V} - \frac{\tau}{1+\tau}\,
{\bfm V}_{\ast}\!\cdot\!\nabla([{\bfm b}\!\cdot\!{\bfm V}]\,{\bfm b})\right]
&=&{\bfm J}\times{\bfm B} - \nabla P \nonumber\\[0.5ex]
&&+ \mu_i\,\nabla^2{\bfm V}_i
+ \mu_e\,\nabla^2{\bfm V}_e,\\[0.5ex]
\left(\frac{\partial}{\partial t} + {\bfm V}\!\cdot\!\nabla\right)\!P
&=& - \Gamma\,P\,\nabla\!\cdot\!{\bfm V} + \kappa\,\nabla^2 P.
\end{eqnarray}
Here, ${\bfm E}$ is the electric field, ${\bfm B}$ the magnetic field,
${\bfm J}$ the electric current density,
${\bfm V}$ the plasma {\em guiding-center} velocity, $P$ the total plasma pressure,
$e$ the magnitude of the electron charge, $m_i$ the
ion mass, $\eta$ the (uniform) plasma resistivity, $\mu_e$
the (uniform) electron viscosity, $\mu_i$ the (uniform) ion viscosity, $\kappa$ the (uniform) plasma thermal
conductivity, and $\Gamma=5/3$ the plasma ratio of specific heats. Furthermore,
${\bfm b} = {\bfm B}/B$, and ${\bfm V}_\ast = {\bfm b}\times\nabla P/
e\,n_0\,B$. The above equations take into account the anisotropic ion gyroviscous tensor,
but neglect electron inertia. Our
system of equations is completed by Maxwell's equations:
$\nabla\!\cdot\!{\bfm B} = 0$, $\nabla\times {\bfm E} = -\partial{\bfm B}/
\partial t$, and $\nabla\times{\bfm B}=\mu_0\,{\bfm J}$. 
Note that the transport coefficients, $\mu_i$, $\mu_e$, and $\kappa$, 
appearing in the above equations, are {\em phenomenological}\/ in nature, and are supposed to represent
the anomalous {\em diffusive}\/  transport of energy and momentum across magnetic
flux-surfaces due to small-scale plasma turbulence

\subsection{Normalized equations}\label{norm}
Let $\hat{\nabla} = a\,\nabla$, $\hat{t} = t/(a/V_a)$,
$\hat{\bfm B} = {\bfm B} /B_a$, $\hat{\bfm E}= {\bfm E}/(B_a\,V_a)$,
$\hat{\bfm J} = {\bfm J}/(B_a/\mu_0\,a)$, $\hat{\bfm V} = {\bfm V}/V_a$,
$\hat{P} = P/(B_a^{\,2}/\mu_0)$, $\hat{\eta} = \eta/(\mu_0\,V_a\,a)$,
$\hat{\mu}_{i,e} = \mu_{i,e}/(n_0\,m_i\,V_a\,a)$, 
$\hat{\kappa} = \kappa/(V_a\,a)$, where $V_a = B_a/\sqrt{\mu_0\,
n_0\,m_i}$. Here, $a$ is a convenient scale length, and $B_a$
a convenient scale magnetic field-strength.

Neglecting hats, our normalized two-fluid equations take the form:
\begin{eqnarray}
{\bfm E} + {\bfm V}\times{\bfm B} +d_i\!
\left(\nabla \,P - \frac{\tau}{1+\tau}\,({\bfm b}\cdot\nabla P)\,{\bfm b}
-{\bfm J}\times{\bfm B} - \mu_e\,\nabla^2{\bfm V}_e\right)
&=&\eta\,\!\left({\bfm J} - \frac{3}{2}\,\frac{\tau}{1+\tau}\,
{\bfm V}_\ast\right),\label{ohm}\\[0.5ex]
\left(
\frac{\partial}{\partial t} + {\bfm V}\!\cdot\!\nabla + \frac{d_i\,\tau}{1+\tau}\,
{\bfm V}_\ast\!\cdot\!\nabla\right)\!{\bfm V} - \frac{d_i\,\tau}{1+\tau}\,
{\bfm V}_{\ast}\!\cdot\!\nabla([{\bfm b}\!\cdot\!{\bfm V}]\,{\bfm b})
&=&{\bfm J}\times{\bfm B} - \nabla P \nonumber\\[0.5ex]
&&+\mu_i\,\nabla^2{\bfm V}_i
+ \mu_e\,\nabla^2{\bfm V}_e,\label{eom}\\[0.5ex]
\left(\frac{\partial}{\partial t} + {\bfm V}\!\cdot\!\nabla\right)\!P
&=& - \Gamma\,P\,\nabla\!\cdot\!{\bfm V} + \kappa\,\nabla^2 P.
\label{energy}
\end{eqnarray}
Here,  ${\bfm V}_\ast =
{\bfm b}\times\nabla P/B$, and
$d_i = (m_i/n_0\,e^2\,\mu_0)^{1/2}/a$ is the normalized
collisionless ion skin-depth. Maxwell's equations are written:
$\nabla\!\cdot\!{\bfm B} = 0$, $\nabla\times {\bfm E} = -\partial{\bfm B}/
\partial t$, and $\nabla\times{\bfm B}={\bfm J}$. 

\subsection{2-D assumption}
Let us make the simplifying assumption that there is no
variation of quantities in  the $z$-direction: {\em i.e.}, $\partial/\partial z\equiv 0$. It
immediately follows that ${\bfm B} = \nabla\psi\times \hat{\bfm z}+ B_z\,\hat{\bfm z}$,
and $E_z = -\partial\psi/\partial t$.

\subsection{Reduction process}
Let us adopt the following ordering, which is designed to decouple the
compressional Alfv\'{e}n wave from all the other waves in the system:
\begin{eqnarray}
P &=& P_0 + B_0\,p_1+p_2,\\[0.5ex]
B_z &=& B_0 + b_z.
\end{eqnarray}
Here, $P_0$ and $B_0$ are {\em uniform and constant}, and
\begin{equation}
P_0\gg B_0\gg 1.
\end{equation}
Furthermore, $p_1$, $p_2$, $b_z$, $\psi$, ${\bfm V}$, $\nabla$, and $\partial/\partial t$
are all assumed to be $O(1)$, and ${\nabla}\!\cdot\!{\bfm V}$
is assumed to be much less than $O(1)$. 

Now, to lowest order, the $z$-component of Ohm's law, Eq.~(\ref{ohm}), gives
\begin{equation}
\left(\frac{\partial}{\partial t} + {\bfm V}\!\cdot\!\nabla\right)
\!\psi = -d_i\,[b_z + \tau\,p_1/(1+\tau), \psi] + \eta\,\nabla^2\psi
- d_i\,\mu_e\,\nabla^2(V_z + d_i\,\nabla^2\psi).
\end{equation}
Here, $[A, B]\equiv \nabla A\times\nabla B\!\cdot\!\hat{\bfm z}.$
Likewise, the $z$-component of the curl of Eq.~(\ref{ohm}) reduces
to
\begin{eqnarray}
\left(\frac{\partial}{\partial t} + {\bfm V}\!\cdot\!\nabla\right)
\!b_z&=& [V_z+ d_i\,\nabla^2\psi,\psi] - B_0\,\nabla\!\cdot\!{\bfm V}
+ \eta\,\nabla^2\!\left(b_z + \frac{3}{2}\frac{\tau}{1+\tau}\,p_1\right)\nonumber\\[0.5ex]
&&+d_i\,\mu_e\,\nabla^2\!\left[U- d_i\,\nabla^2\left(b_z+ \frac{\tau}{1+\tau}
\,p_1\right)\right].\label{e1}
\end{eqnarray}
Here, $U = - \nabla\times{\bfm V}\!\cdot\!\hat{\bfm z}$.

To lowest order, the equation of motion, Eq.~(\ref{eom}), implies that
\begin{equation}
p_1 \simeq-b_z. \label{e3}
\end{equation}
Furthermore, the $z$-component of this equation yields
\begin{equation}
\left(\frac{\partial}{\partial t} + {\bfm V}\!\cdot\!\nabla\right)
\!V_z = [b_z, \psi] + \mu_i\,\nabla^2 V_z + \mu_e\,\nabla^2(V_z
+ d_i\,\nabla^2\psi),
\end{equation}
whereas the $z$-component of its curl reduces to
\begin{eqnarray}
\left(\frac{\partial}{\partial t} + {\bfm V}\!\cdot\!\nabla\right)
\!U &=&- \frac{d_i}{2}\frac{\tau}{1+\tau}
\left\{\nabla^2[\phi,b_z] + [U,b_z]+ [\nabla^2 b_z,\phi]\right\} + [\nabla^2\psi,\psi] \nonumber\\[0.5ex]
&&+ \mu_i\,\nabla^2\left(U + \frac{d_i\,\tau}{1+\tau}
\nabla^2 b_z\right) + \mu_e\,\nabla^2\left(U - \frac{d_i}{1+\tau}\,
\nabla^2 b_z\right).
\end{eqnarray}

Finally, to  lowest order, the energy equation, Eq.~(\ref{energy}), gives
\begin{equation}
\left(\frac{\partial}{\partial t} + {\bfm V}\!\cdot\!\nabla\right)\!p_1
= - \frac{\Gamma\,P_0}{B_0}\,\nabla\!\cdot\!{\bfm V} + \kappa\,
\nabla^2 p_1.\label{e2}
\end{equation}

Eliminating $\nabla\!\cdot\!{\bfm V}$ between Eqs.~(\ref{e1}) and
(\ref{e2}), making use of Eq.~(\ref{e3}), we obtain
\begin{eqnarray}
c_\beta^{-2}\left(\frac{\partial}{\partial t} + {\bfm V}\!\cdot\!\nabla\right)
\!b_z&=& [V_z+ d_i\,\nabla^2\psi,\psi] 
+\left[ \eta\left(1-\frac{3}{2}\frac{\tau}{1+\tau}\right)+\frac{\kappa}{\beta}\right]\nabla^2
b_z\nonumber\\[0.5ex]
&&+d_i\,\mu_e\,\nabla^2\!\left(U- \frac{d_i}{1+\tau}\,\nabla^2 b_z\right).
\end{eqnarray}
Here, $\beta = \Gamma\,P_0/B_0^{\,2}$ is ($\Gamma$ times) the plasma beta calculated with the ``guide-field'', $B_0$, and $c_\beta=
\sqrt{\beta/(1+\beta)}$. Note that our ordering scheme does not
constrain $\beta$ to be either much less than or much greater than unity.

Equation~(\ref{e2}) implies that $\nabla\!\cdot\!{\bfm V} \sim O(B_0^{-1})$: {\em i.e.},  that the flow is {\em almost} incompressible. Hence, to
lowest order, we can write
\begin{equation}
{\bfm V} = \nabla\phi\times \hat{\bfm z} + V_z\,\hat{\bfm z}.
\end{equation}

\subsection{Final equations}
Let $d_\beta = c_\beta\,d_i/\sqrt{1+\tau}$, $Z=b_z/c_\beta\,\sqrt{1+\tau}$,
and $\bar{V}_z= V_z/\sqrt{1+\tau}$. Neglecting the bar over $\bar{V}_z$,
our final set of reduced, 2-D, two-fluid, drift-MHD equations takes the form:
\begin{eqnarray}
\frac{\partial\psi}{\partial t}&=& [\phi - d_\beta\,Z,\psi]
+ \eta\,J - \frac{\mu_e\,d_\beta\,(1+\tau)}{c_\beta}\,
\nabla^2[V_z + (d_\beta/c_\beta)\,J],\label{efa}\\[0.5ex]
\frac{\partial Z}{\partial t} &=& [\phi,Z] + c_\beta\,[V_z+
(d_\beta/c_\beta)\,J,\psi] + c_\beta^{\,2}\left[\eta\!\left(1-\frac{3}{2}\frac{\tau}
{1+\tau}\right) + \frac{\kappa}{\beta}\right]Y\nonumber\\[0.5ex]&& + \mu_e\,d_\beta\,\nabla^2(U-d_\beta\,Y),\\[0.5ex]
\frac{\partial U}{\partial t} &=& [\phi,U]
- \frac{d_\beta\,\tau}{2} \left\{\nabla^2[\phi,Z]+[U,Z]+[Y,\phi]\right\}
+ [J,\psi]+\mu_i\,\nabla^2(U + d_\beta\,\tau\,Y) \nonumber\\[0.5ex]&&+ 
\mu_e\,\nabla^2(U-d_\beta\,Y),\\[0.5ex]
\frac{\partial V_z}{\partial t} &=& [\phi,V_z]+c_\beta\,[Z,\psi]
+\mu_i\,\nabla^2 V_z + \mu_e \,\nabla^2[V_z+ (d_\beta/c_\beta)\,J].\label{efd}
\end{eqnarray}
Here, $U=\nabla^2\phi$, $J=\nabla^2\psi$, and $Y=\nabla^2 Z$.
The four fields which are evolved in the above equations are the magnetic flux-function, $\psi$,
the (normalized) perturbed $z$-directed magnetic field, $Z$ ($=b_z/c_\beta\,\sqrt{1+\tau}$),  the
$z$-directed guiding-center vorticity, $U$, and the (normalized) $z$-directed guiding-center (and ion) fluid velocity, $V_z$ ($={\bfm V}\!\cdot\!\hat{\bfm z}/\sqrt{1+\tau}$). The (normalized) $z$-directed electron
fluid velocity is $V_z+(d_\beta/c_\beta)\,J$.
The quantity $\phi$ is the guiding-center stream-function. The ion stream-function takes the form $\phi_i=\phi+d_\beta\,\tau\,Z$,
whereas the electron stream-function is written $\phi_e =\phi-d_\beta\,Z$.
The above equations are ``reduced" in the sense that they
do not
contain the compressible Alfv\'{e}n wave. However, they do contain the
shear-Alfv\'{e}n wave, the magnetoacoustic wave, the whistler wave,
and the kinetic-Alfv\'{e}n wave. Our equations are similar to the
``four-field" equations of Hazeltine, Kotschenreuther, and Morrison,\cite{haz1}
except that they are not limited to small values of $\beta$.

\section{Island physics}
\subsection{Introduction}
The aim of this section is to derive expressions determining the 
phase-velocity and width of an {\em isolated} magnetic island (representing
the final, nonlinear stage of a tearing instability) from the
previously derived set of reduced, 2-D, two-fluid, drift-MHD equations.

Consider a {\em slab} plasma which is periodic in the $y$-direction
with periodicity length $l$. Let the system be
{\em symmetric} about $x=0$: {\em i.e.}, $\psi(-x,y,t)=\psi(x,y,t)$, 
$Z(-x,y,t)=-Z(x,y,t)$,
$\phi(-x,y,t)=-\phi(x,y,t)$,  and $V_z(-x,y,t)=V_z(x,y,t)$. 
Consider a quasi-static, constant-$\psi$ magnetic island, centered on $x=0$.
It is convenient to transform to the
{\em island rest-frame}, in which $\partial/\partial t\simeq 0$.
Suppose that the island is embedded in a plasma with {\em uniform}
(but different) $y$-directed ion and electron fluid  velocities. We
are searching for an island solution in which the ion/electron fluid velocities
asymptote to these uniform velocities far from the island separatrix.

\subsection{Island geometry}
In the immediate vicinity of the island, we can write
\begin{equation}
\psi(x,\theta,t) = -\frac{x^2}{2} + {\mit\Psi}(t)\,\cos\theta,
\end{equation}
where $\theta = k\,y$, $k=2\pi/l$, and ${\mit\Psi}(t)>0$ is the reconnected magnetic
flux (which is assumed to have a very weak time dependence). As is well-known, the above expression for $\psi$ describes
a ``cat's eye" magnetic island of full-width (in the $x$-direction)
$W=4\,w$, where $w=\sqrt{\mit\Psi}$. 
The region inside the magnetic separatrix corresponds to $\psi>-{\mit\Psi}$, the region outside the separatrix corresponds to $\psi<-{\mit\Psi}$, and the separatrix itself
corresponds to $\psi=-{\mit\Psi}$. The island O- and X-points are located
at $(x,\theta) = (0,0)$, and $(x,\theta)=(0,\pi)$, respectively.

It is helpful to define a flux-surface average operator:
\begin{equation}
\langle f(s,\psi,\theta) \rangle= \oint \frac{f(s,\psi,\theta)}{|x|}\,\frac{d\theta}{2\pi}
\end{equation}
for $\psi\leq -{\mit\Psi}$, and
\begin{equation}
\langle f(s,\psi,\theta) \rangle = \int_{-\theta_0}^{\theta_0}
\frac{f(s,\psi,\theta)+f(-s,\psi,\theta)}{2\,|x|}\,\frac{d\theta}{2\pi}
\end{equation}
for $\psi> -{\mit\Psi}$. Here, $s={\rm sgn}(x)$, and $x(s,\psi,\theta_0)=0$
(with $\pi>\theta_0>0$).
The most important property of this operator is that
\begin{equation}
\langle [A,\psi]\rangle \equiv 0,
\end{equation}
for any field $A(s,\psi,\theta)$. 

\subsection{Island equations}
The equations governing the quasi-static island [which follow from
Eqs.~(\ref{efa})--(\ref{efd})] are:
\begin{eqnarray}
\frac{d{\mit\Psi}}{dt}\,\cos\theta&=& [\phi - d_\beta\,Z,\psi]
+ \eta\,\delta J - \frac{\mu_e\,d_\beta\,(1+\tau)}{c_\beta}\,
\nabla^2[V_z + (d_\beta/c_\beta)\,\delta J]\label{ea},\\[0.5ex]
0&=& [\phi,Z] + c_\beta\,[V_z+
(d_\beta/c_\beta)\,\delta J,\psi] + c_\beta^{\,2}\,D\,Y + \mu_e\,d_\beta\,\nabla^2(U-d_\beta\,Y),\label{eb}\\[0.5ex]
0 &=& [\phi,U]
- \frac{d_\beta\,\tau}{2} \left\{\nabla^2[\phi,Z]+[U,Z]+[Y,\phi]\right\}
+ [\delta J,\psi]+\mu_i\,\nabla^2(U + d_\beta\,\tau\,Y) \nonumber\\[0.5ex]&&+ 
\mu_e\,\nabla^2(U-d_\beta\,Y),\label{ec}\\[0.5ex]
0&=& [\phi,V_z]+c_\beta\,[Z,\psi]
+\mu_i\,\nabla^2 V_z + \mu_e \,\nabla^2[V_z+ (d_\beta/c_\beta)\,\delta J],\label{ed}
\end{eqnarray}
where $\delta J=1+\nabla^2\psi$ (the 1 represents an externally applied, inductive
electric field maintaining the equilibrium plasma current), $Y=\nabla^2 Z$,  $U=\nabla^2\phi$,
and
\begin{equation}
D = \eta\!\left(1-\frac{3}{2}\frac{\tau}
{1+\tau}\right) + \frac{\kappa}{\beta}.
\end{equation}

\subsection{Ordering scheme}\label{order}
We adopt the following ordering of terms appearing in Eqs.~(\ref{ea})--(\ref{ed}):
$\psi = \psi^{(0)}$, $\phi=\phi^{(1)}(s,\psi) + \phi^{(3)}(s,\psi,\theta)$,
$Z= Z^{(1)}(s,\psi)+Z^{(3)}(s,\psi,\theta)$, $V_z=V_z^{(2)}(s,\psi,\theta)$, $\delta J = 
\delta J^{(2)}(s,\psi,\theta)$. Moreover, $\nabla=\nabla^{(0)}$, $\tau=\tau^{(0)}$,
$c_\beta = c_\beta^{(0)}$, $d_\beta =d_\beta^{(0)}$, $\mu_{i,e}=
\mu_{i,e}^{(2)}$, $\kappa=\kappa^{(2)}$, $\eta =\eta^{(2)}$,
and $d{\mit\Psi}/dt=d{\mit\Psi}^{(4)}/dt$. 
Here, the superscript $^{(i)}$ indicated an $i$th order quantity.
This ordering, which is completely self-consistent,  implies weak ({\em i.e.},
strongly sub-Alfv\'{e}nic and sub-magnetoacoustic) diamagnetic  flows, and very long ({\em i.e.}, very much
longer than the Alfv\'{e}n time) transport evolution time-scales.
According to our  scheme,  both $Z$ and $\phi$ are {\em flux-surface
functions}, to lowest order. In other words, the lowest order electron and ion stream-functions, $\phi_e
=\phi - d_\beta \,Z$ and $\phi_i=\phi+d_\beta\,\tau\,Z$, respectively, are flux-surface
functions. 

To lowest and next lowest orders, Eqs.~(\ref{ea})--(\ref{ed}) yield:
\begin{eqnarray}
\frac{d{\mit\Psi}^{(4)}}{dt}\,\cos\theta&=& [\phi^{(3)} - d_\beta\,Z^{(3)},\psi]
+ \eta^{(2)}\,\delta J^{(2)} - \frac{\mu_e^{(2)}\,d_\beta\,(1+\tau)}{c_\beta}\,
\nabla^2[V_z^{(2)} + (d_\beta/c_\beta)\,\delta J^{(2)}]\label{eva},\\[0.5ex]
0&=& c_\beta\,[V_z^{(2)}+
(d_\beta/c_\beta)\,\delta J^{(2)},\psi] +c_\beta^{\,2}\,D^{(2)}\,Y^{(1)} + \mu_e^{(2)}\,d_\beta\,\nabla^2(U^{(1)}-d_\beta\,Y^{(1)}),\label{evb}\\[0.5ex]
0 &=&- M^{(1)}\,[U^{(1)},\psi]
- \frac{d_\beta\,\tau}{2} \left\{L^{(1)}\,[U^{(1)},\psi]+M^{(1)}\,[Y^{(1)},\psi]\right\}
+ [\delta J^{(2)},\psi]\nonumber\\[0.5ex]&&+\mu_i^{(2)}\,\nabla^2(U^{(1)} + d_\beta\,\tau\,Y^{(1)})+ 
\mu_e^{(2)}\,\nabla^2(U^{(1)}-d_\beta\,Y^{(1)}),\label{evc}\\[0.5ex]
0&=&- M^{(1)}\,[V_z^{(2)},\psi]+c_\beta\,[Z^{(3)},\psi]
+\mu_i^{(2)}\,\nabla^2 V_z^{(2)} + \mu_e^{(2)} \,\nabla^2[V_z^{(2)}+ (d_\beta/c_\beta)\,\delta J^{(2)}],\label{evd}
\end{eqnarray}
where $Y^{(1)}=\nabla^2 Z^{(1)}$,  $U^{(1)}=\nabla^2\phi^{(1)}$,
$M^{(1)}(s,\psi)=d\phi^{(1)}/d\psi$, and $L^{(1)}(s,\psi)=dZ^{(1)}/d\psi$.
Here, we have neglected the superscripts on zeroth order quantities,
for the sake of clarity. 
In the following, we shall neglect all superscripts,
except for those on $\phi^{(3)}$ and $Z^{(3)}$, for ease of notation.

\subsection{Boundary conditions}
 It is easily demonstrated that the $y$-components of the (lowest order) electron and ion  fluid velocities (in the island
rest frame)
take the form $V_{e\,y} = x\,(M-d_\beta\,L)$ and $V_{i\,y}=x\,(M
+d_\beta\,\tau \,L)$, respectively. Incidentally, since $V_{e\,y}$ and $V_{i\,y}$
are  even functions of $x$, it follows that $M(s,\psi)$ and $L(s,\psi)$ are
odd functions. We immediately conclude that $M(s,\psi)$ and $L(s,\psi)$ are both
{\em zero} inside the island separatrix (since it is impossible to have a non-zero
odd flux-surface function in this region). Now, we are searching for
island solutions for which $x\,M\rightarrow M_0$ and
$x\,L\rightarrow L_0$ as $|x|/w\rightarrow \infty$. In other words,
we desire solutions which match to an unperturbed plasma far from the island.
If $V_{e\,y}^{(0)}$ and $V_{i\,y}^{(0)}$ are the unperturbed $y$-directed
electron and ion fluid  velocities in the {\em lab.\ frame}, then
$V_{e\,y}^{(0)}-V= M_0-d_\beta\,L_0$ and $V_{i\,y}^{(0)}-V=
M_0+d_\beta\,\tau\,L_0$, where $V$ is the island phase-velocity
in the {\em lab.\ frame}. It follows that $L_0 = (V_{i\,y}^{(0)}-
V_{e\,y}^{(0)}) / d_\beta\, (1+\tau)$ and $M_0 = V_{EB\,y}^{(0)} - V$,
where $V_{EB\,y}^{(0)}=(V_{i\,y}^{(0)}+\tau\,V_{e\,y}^{(0)})/(1+\tau)$ is the unperturbed plasma ${\bfm E}\times {\bfm B}$
velocity in the lab.\ frame. Hence, determining the island 
phase-velocity is equivalent to determining the value of $M_0$.

\subsection{Determination of  flow profiles}
Flux-surface averaging Eqs.~(\ref{evb}) and (\ref{evc}), we
obtain
\begin{equation}\label{ex}
\langle \nabla^2 U\rangle +\frac{d_\beta\,(\mu_i\,\tau-\mu_e)}
{(\mu_i+\mu_e)}\,\langle\nabla^2 Y\rangle=0,
\end{equation}
and
\begin{equation}\label{ey}
\delta^2\,w^2\,\langle \nabla^2 Y\rangle -\langle Y\rangle =0,
\end{equation}
where
\begin{equation}
\delta=\frac{d_i}{w\,\sqrt{D}}\,\sqrt{\frac{\mu_i\,\mu_e}{\mu_i+
\mu_e}}.
\end{equation}

Assuming that the island is ``thin'' ({\em i.e.}, $w\ll l$), we can
write $\nabla^2\simeq\partial^2/\partial x^2$. 
Hence, Eqs.~(\ref{ex}) and (\ref{ey}) yield
\begin{equation}\label{edd}
M(s,\psi) =  - \frac{d_\beta\,(\mu_i\,\tau-\mu_e)}
{(\mu_i+\mu_e)}\,L(s,\psi) + F(s,\psi),
\end{equation}
where
\begin{equation}\label{ez}
\frac{d}{d\psi}\!\left[\frac{d}{d\psi}\!\left(\delta^2\,w^2\,\langle x^4\rangle\,\frac{d L}{d
\psi}\right)-\langle x^2\rangle\,L\right] =0,
\end{equation}
and
\begin{equation}\label{ezz}
\frac{d^2}{d\psi^2}\!\left(\langle x^4\rangle\,\frac{d F}{d
\psi}\right)=0.
\end{equation}

We can integrate Eq.~(\ref{ez}) once to give
\begin{equation}\label{exx}
\delta^2\,w^2\,\frac{d}{d\psi}\!\left(\langle x^4\rangle\,\frac{d L}{d
\psi}\right)-\langle x^2\rangle\,L =-s\,L_0.
\end{equation}
We can solve Eq.~(\ref{ezz}), subject to the constraints that $F$ be
continuous, $F=0$ inside the separatrix, and $F\rightarrow s\,F_0$
as $|x|/w\rightarrow\infty$, to give
\begin{equation}\label{ef}
F(s,\psi)= s\,F_0\,\left.\int_{-{\mit\Psi}}^\psi\frac{d\psi}{\langle x^4\rangle}
\right/ \int_{-{\mit\Psi}}^{-\infty}\frac{d\psi}{\langle x^4\rangle}
\end{equation}
outside the separatrix. Note that $x\,F\rightarrow |x|\,F_0$ as
$|x|/w\rightarrow\infty$.

In order to solve Eq.~(\ref{exx}), we write
$\hat{\psi}=-\psi/{\mit\Psi}$, $\langle\!\langle\cdots\rangle\!\rangle=
\langle\cdots\rangle w$, $X=x/w$, and $\hat{L} = L/(L_0/w)$.
It follows that
\begin{equation}
\delta^2\,\frac{d}{d\hat{\psi}}\!\left(\langle\!\langle X^4\rangle\!\rangle\,\frac{d \hat{L}}{d
\hat{\psi}}\right)-\langle\!\langle X^2\rangle\!\rangle\,\hat{L}=-s.
\end{equation}
Suppose that $\delta\ll 1$. In this case, $\hat{L}(s,\hat{\psi})$ takes the value
$s/\langle\!\langle X^2\rangle\!\rangle$ in the region outside the magnetic
separatrix, apart from a thin boundary layer on
the separatrix itself of width $\delta\,w$. In this layer, the function $\hat{L}(s,\hat{\psi})$
makes a smooth transition from its exterior value (which is $s\,\pi/4$
immediately outside the separatrix) to its interior value $0$.
We can write
\begin{equation}
\hat{L}(s,\hat{\psi}) = s\left(\frac{1}{\langle\!\langle X^2\rangle\!\rangle}+l(y)\right),
\end{equation}
where $y = (\hat{\psi}-1)/\delta$. It follows that
\begin{equation}
\frac{d^2 l}{dy^2}-\frac{3}{8}\,l \simeq 0,
\end{equation}
since $\langle\!\langle X^2\rangle\!\rangle_{\hat{\psi}=1} = 4/\pi$,
and $\langle\!\langle X^4\rangle\!\rangle_{\hat{\psi}=1}=32/3\,\pi$.
Hence, the continuous solution to Eq.~(\ref{exx}) which satisfies the appropriate boundary
conditions is
\begin{equation}\label{ldef}
\hat{L}(s,\hat{\psi})=  s\left[\frac{1}{\langle\!\langle X^2\rangle\!\rangle}-\frac{\pi}{4}\,
\exp\!\left(-\sqrt{\frac{3}{8}}\,\frac{\hat{\psi}-1}{\delta}\right)\right]
\end{equation}
in the region outside the separatrix ({\em i.e.}, $\hat{\psi}\geq 1$).
Of course, $\hat{L}(s,\hat{\psi})=0$ in the region inside the separatrix
({\em i.e.}, $\hat{\psi}<1$).

\subsection{Determination of island phase-velocity}
Let $\delta J = \delta J_c + \delta J_s$, where $\delta J_c$ has the
symmetry of $\cos\theta$, whereas $\delta J_s$ has the symmetry of
$\sin\theta$. Now, it is easily
demonstrated that
\begin{equation}
\langle \delta J_s\,\sin\theta\rangle = \frac{1}{k\,{\mit\Psi}}
\langle x\,[\delta J, \psi]\rangle.
\end{equation}
Hence, it follows from Eq.~(\ref{evc}) and (\ref{edd})
that 
\begin{equation}\label{err}
\langle \delta J_s\,\sin\theta\rangle = - \frac{(\mu_i+\mu_e)}{k\,{\mit\Psi}}
\,\frac{d}{d\psi}\!\left(\langle
x^5\rangle\,\frac{d^2 F}{d\psi^2}- 2\,\langle x^3\rangle
\,\frac{d F}{d\psi} - \langle x\rangle F\right).
\end{equation}
Now, for an isolated magnetic island which is not interacting electromagnetically
with any external structure, such as a resistive wall, the net electromagnetic
force acting on the island must be {\em zero}. This constraint translates
to the well-known requirement that
\begin{equation}\label{sin}
\int_{-{\mit\Psi}}^\infty \langle \delta J_s\,\sin\theta\rangle\,d\psi = 0.
\end{equation}
Using Eq.~(\ref{err}), this requirement reduces to the condition
\begin{eqnarray}
\lim_{x/w\rightarrow \infty}\left(\langle
x^5\rangle\,\frac{d^2 F}{d\psi^2}-2\,\langle x^3\rangle
\,\frac{d F}{d\psi} - \langle x\rangle F\right) &\propto&
\lim_{x/w\rightarrow \infty} \left[s\,x^2\,\frac{d}{dx}\!\left(
\frac{1}{x}\,\frac{d(x\,F)}{dx}\right)\right]\nonumber\\[0.5ex]&=& -F_0=0,
\end{eqnarray}
since $x\,F\rightarrow |x|\,F_0$ as $|x|/w\rightarrow \infty$.
Hence, we conclude that $F_0=0$ [{\em i.e.}, $F(\psi)=0$,
everywhere] for an isolated magnetic island. 

\begin{figure}
\epsfysize=2.8in
\centerline{\epsffile{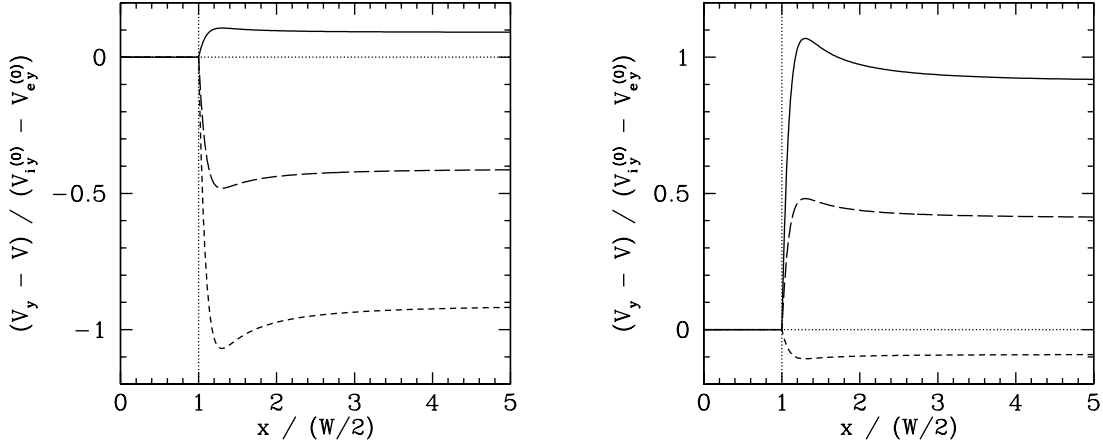}}
\caption{\em Velocity profiles as functions of $x$, at
constant $\theta$, evaluated on a line passing through the island O-point ({\em i.e.},
at $\theta = 0$) in the island rest frame. The O-point lies
at $x=0$. The island separatrix
is indicated by a vertical dotted line. The solid curves show the
normalized ion fluid velocity profile: $(V_{i\,y}-V) / (V_{i\,y}^{(0)}-V_{e\,y}^{(0)})$. The short-dashed curves show the normalized electron
fluid velocity profile:  $(V_{e\,y}-V) / (V_{i\,y}^{(0)}-V_{e\,y}^{(0)})$.
The long-dashed curves show the normalized ${\bfm E}\times{\bfm B}$ velocity profile:  $(V_{EB\,y}-V) / (V_{i\,y}^{(0)}-V_{e\,y}^{(0)})$.
The left-hand panel shows the case of viscous ions: $\mu_e/\mu_i=0.1$,
$\tau =1.$, and $\delta = 0.2$. The right-hand panel shows the case of viscous electrons: $\mu_i/\mu_e=0.1$,
$\tau =1.$, and $\delta = 0.2$. }\label{f1}
\end{figure}

It follows from Eq.~(\ref{edd}) that
\begin{equation}
M(s,\psi) = - \frac{d_\beta\,(\mu_i\,\tau-\mu_e)}
{(\mu_i+\mu_e)}\,L(s,\psi) .
\end{equation}
Hence, $M_0=-[d_\beta\,(\mu_i\,\tau-\mu_e)/(\mu_i+\mu_e)]\,L_0$.
Recalling that $M_0 = V_{EB\,y}^{(0)} - V$, 
$d_\beta\,L_0 = (V_{i\,y}^{(0)}-V_{e\,y}^{(0)})/(1+\tau)$, $V_{i\,y}^{(0)}
=V_{EB\,y}^{(0)} + d_\beta\,\tau\,L_0$, and
$V_{e\,y}^{(0)} = V_{EB\,y}^{(0)}-d_\beta\,L_0$,
we obtain the following expression for the island phase-velocity:
\begin{equation}\label{freq}
V = \frac{\mu_i\,V_{i\,y}^{(0)} + \mu_e\,V_{e\,y}^{(0)}}{\mu_i + \mu_e}.
\end{equation}
In other words, the island phase-velocity is the viscosity
weighted mean of the unperturbed ion and electron fluid  velocities. Hence, if the
ions are far more viscous then the electrons, then the island propagates with
the ion fluid. In this case, the ion fluid velocity profile remains largely unaffected by the
island, but the electron fluid velocity profile is highly sheared just outside
the island separatrix. The opposite is true if the electrons are far more 
viscous than the ions. This is illustrated in Fig.~\ref{f1}.

We have now fully specified the ion and electron stream-functions,
$\phi_i$ and $\phi_e$, respectively, in the island rest frame. In fact, $\phi_i=0$ inside the separatrix,
and 
\begin{equation}
\frac{d\phi_i(s,\hat{\psi})}{d\psi} = (V_{i\,y}^{(0)}-V_{e\,y}^{(0)})\,\frac{\mu_i}{\mu_i+\mu_e}\,\frac{\hat{L}(s,\hat{\psi})}{w}
\end{equation}
outside the separatrix, where the function $\hat{L}(s,\hat{\psi})$ is specified in Eq.~(\ref{ldef}).
Likewise, $\phi_e$ is zero inside the separatrix, and
\begin{equation}
\frac{d\phi_e(s,\hat{\psi})}{d\psi} = -(V_{i\,y}^{(0)}-V_{e\,y}^{(0)})\,\frac{\mu_e}{\mu_i+\mu_e}\,\frac{\hat{L}(s,\hat{\psi})}{w}
\end{equation}
outside the separatrix. Note that the stream-functions and their first derivatives
are everywhere continuous, which implies that the ion and electron fluid  velocities
are everywhere continuous.

\subsection{Determination of  ion polarization correction}
It follows from Eq.~(\ref{evc}) that
\begin{equation}\label{eii}
\delta J_c = \frac{(V-V_{EB\,y}^{(0)})\,(V-V_{i\,y}^{(0)})}{2}
\left(x^2-\frac{\langle x^2\rangle}{\langle 1\rangle}\right)
\frac{d}{d\psi}\!\left(\frac{1}{\langle x^2\rangle^2}\right)H(\hat{\psi}-1) + I(s,\psi),
\end{equation}
where $I(s,\psi)$ is as yet undetermined. The function $H(\vartheta)$ is zero for $\vartheta<0$,
and unity for $\vartheta \geq 0$.
Here, we have made use of the
fact that outside the separatrix $L(s,\psi)\simeq s\,L_0/\langle x^2\rangle$,
and $M(s,\psi)\simeq s\,M_0/\langle x^2\rangle$, apart from a thin
boundary layer on the separatrix itself. It turns out that we do not need to
resolve this boundary layer in order to calculate the total ion polarization current. However, we do have to include the net current flowing in this 
layer in our calculation of the total current.\cite{wal,rf}
Flux-surface averaging Eqs.~(\ref{eva}) and (\ref{evd}), we obtain
\begin{equation}\label{edj}
\epsilon^2\,w^2\,\langle \nabla^2\delta J_c\rangle -\langle\delta J_c\rangle
= -\eta^{-1}\,\frac{d{\mit\Psi}}{dt}\,\langle \cos\theta\rangle,
\end{equation}
where
\begin{equation}
\epsilon = \frac{d_i}{w\,\sqrt{\eta}}\,\sqrt{\frac{\mu_i\,\mu_e}{\mu_i
+\mu_e}}.
\end{equation}

Equation~(\ref{edj}) implies that
\begin{equation}\label{ei}
\langle \delta J_c\rangle \simeq \eta^{-1}\,\frac{d{\mit\Psi}}{dt}\,\langle \cos\theta\rangle,
\end{equation}
apart from in a thin boundary layer on the separatrix of width $\epsilon\,w$.
Here, we are assuming that $\epsilon\ll 1$. It is easily demonstrated 
that the deviation of $\langle \delta J_c\rangle$ in the boundary layer 
from the value given
in Eq.~(\ref{ei})  makes a negligible contribution
to the total ion polarization current. Hence, we shall treat Eq.~(\ref{ei})
as if it applied everywhere.

Equations (\ref{eii}) and (\ref{ei}) give
\begin{equation}\label{eqq}
\delta J_c = \frac{(V-V_{EB\,y}^{(0)})\,(V-V_{i\,y}^{(0)})}{2}
\left(x^2-\frac{\langle x^2\rangle}{\langle 1\rangle}\right)
\frac{d}{d\psi}\!\left(\frac{1}{\langle x^2\rangle^2}\right) H(\hat{\psi}-1)+\eta^{-1}\,\frac{d{\mit\Psi}}{dt}\, \frac{\langle\cos\theta\rangle}{\langle 1\rangle}.
\end{equation}
Note that this current profile contains no discontinuities or singularities.

The island width evolution equation is obtained by asymptotic matching
to the region far from the island.\cite{ruth} In fact,
\begin{equation}\label{epp}
\Delta'\,{\mit\Psi} = -4\int_{{\mit\Psi}}^{-\infty} \langle
\delta J_c\,\cos\theta\rangle\,d\psi,
\end{equation}
where $\Delta'$ is the conventional tearing stability index.\cite{fkr}
It follows from Eqs.~(\ref{eqq}) and (\ref{epp}) that
\begin{eqnarray}
\Delta' &= & -\frac{(V-V_{EB\,y}^{(0)})\,(V-V_{i\,y}^{(0)})}{w^3}
\int_{+1}^\infty\left(\langle\!\langle X^4\rangle\!\rangle
-\frac{\langle\!\langle X^2\rangle\!\rangle^2}{\langle\!\langle
1 \rangle\!\rangle}\right)\frac{d}{d\hat{\psi}}\!\left(
\frac{1}{\langle\!\langle X^2\rangle\!\rangle^2}\right)\,d\hat{\psi}\nonumber\\[0.5ex]&&
+\frac{8}{\eta}\frac{dw}{dt}\int_{-1}^\infty
\frac{\langle\!\langle\cos\theta\rangle\!\rangle^2}{\langle\!\langle 1
\rangle\!\rangle}\,d\hat{\psi}.
\end{eqnarray}
Performing the flux-surface integrals, whose values are well-known,\cite{rf}
we obtain the following island width evolution equation:
\begin{equation}\label{wo}
\frac{0.823}{\eta}\,\frac{dW}{dt} = \Delta'
+ 1.38\,\frac{(V-V_{EB\,y}^{(0)})\,(V-V_{i\,y}^{(0)})}{(W/4)^3}.
\end{equation}
Here, $W=4\,w$ is the full island width. The ion polarization current
term (the second term on the r.h.s.)\ is stabilizing when the
island phase-velocity, $V$, lies between the unperturbed local ${\bfm E}\times {\bfm V}$
velocity, $V_{EB\,y}^{(0)}$, and the unperturbed local velocity of the ion fluid,
$V_{i\,y}^{(0)}$.\cite{con1}

\section{Summary and disscusion}
A set of reduced, 2-D, two-fluid, drift-MHD
equations is developed. This set of equations takes into account both electron and ion diamagnetism (including the contribution of the ion gyroviscous tensor),
as well as the Hall effect and parallel electron compressibility,
but neglects electron inertia and the compressible Alfv\'{e}n wave. For the
sake of simplicity, the plasma density is assumed to be uniform, and
the ion and electron temperatures constant multiples of one another.
However, these constraints could easily be relaxed.

Using
our equations, we have derived a complete and self-consistent solution
for an isolated magnetic island propagating
through a slab plasma with {\em uniform} but {\em different}\/ ion and electron fluid 
velocities.  Our solution is valid
provided that the ordering scheme described
in Sect.~\ref{order} holds good, and the island width $W$ is sufficiently large
that 
\begin{equation}
W \gg  \frac{d_i}{\sqrt{D}}\,\sqrt{\frac{\mu_i\,\mu_e}{\mu_i+\mu_e}}
\end{equation}
({\em i.e.,} $\delta\ll 1$), and
\begin{equation}
W\gg  \frac{d_i}{\sqrt{\eta}}\,\sqrt{\frac{\mu_i\,\mu_e}{\mu_i+\mu_e}}
\end{equation}
({\em i.e.}, $\epsilon\ll 1$). 

Note that the ordering scheme
described in Sect.~\ref{order} implies that $\omega_\ast \ll
k_\parallel\,c_s, k_\parallel\,v_A$ where $\omega_\ast$ is a
typical diamagnetic frequency,   $c_s$ the sound speed,
and $v_A$ the shear-Alfv\'{e}n speed. Here, $k_\parallel$ must be evaluated
at the {\em edge of the island}.
This scheme differs from that adopted in Ref.~\onlinecite{con1},
for which $k_\parallel\,c_s\ll \omega_\ast$.
 It turns out that our ordering scheme permits a much
less complicated calculation of the flow profiles around the island
than that described in Ref.~\onlinecite{con1}. 

Within our solution, the ion and electron fluid velocity profiles are {\em uniquely
determined}\/ in the vicinity of the island [see Fig.~\ref{f1}]. These profiles are everywhere
{\em continuous} and asymptote to the unperturbed fluid velocities far from
the island. Incidentally, the inclusion of {\em electron viscosity}\/ in both the
Ohm's law and the plasma equation of motion is key to the determination
of {\em continuous} velocity profiles.\cite{mik}

The island phase-velocity is uniquely specified by the
condition that there be zero net electromagnetic
force acting on the island [see Eq.~(\ref{freq})].
It turns out that the phase-velocity is the viscosity
weighted mean of the unperturbed ion and electron fluid velocities.
In this paper, we have adopted phenomenological {\em diffusive}
ion and electron viscosity operators, which are supposed to
represent  anomalous perpendicular momentum transport due to
small-scale plasma turbulence. 

The ion polarization current correction to the Rutherford island
width evolution equation is found to be stabilizing when the
island phase-velocity lies between the unperturbed ion fluid  velocity
and the unperturbed ${\bfm E}\times{\bfm B}$ velocity [see Eq.~(\ref{wo})].\cite{con1} It follows, from our result for the island phase-velocity,
that the polarization term is {\em stabilizing}\/ when the anomalous perpendicular ion
viscosity {\em significantly exceeds} the anomalous perpendicular
electron viscosity [see Fig.~\ref{f1}, left panel].
Conversely, the polarization term is destabilizing when the
electron viscosity significantly exceeds the ion viscosity [see Fig.~\ref{f1}, right panel].\cite{chris}
Note, however, that in order for the electron viscosity to
exceed the ion viscosity, the electron momentum confinement
time would need to be at least a {\em mass ratio smaller}\/ than the ion
momentum confinement time, which does not seem very probable.
Hence, we conclude that under normal circumstances the polarization
term is {\em stabilizing}.

\subsection*{Acknowledgments}
This research was inspired by a remote seminar given by Chris Hegna,
as part of the Office of Fusion Energy Sciences' Theory Seminar Series, and
 was funded by
the U.S.\ Department of Energy under contract DE-FG05-96ER-54346.

\end{document}